\documentclass[a4paper,11pt]{article}
\usepackage{pos}
\usepackage{mathtools}

\graphicspath{{./figures/}}
\title{Thermal QCD phase transition\\with dynamical chiral fermions}
\ShortTitle{Thermal QCD phase transition with dynamical chiral fermions}

\author[a,b,c,d]{Z.~Fodor}
\author*[b]{A.Yu.~Kotov}
\author[a,b]{K.~Szabo}

\affiliation[a]{Department of Physics, University of Wuppertal, D-42119 Wuppertal, Germany}

\affiliation[b]{J\"ulich Supercomputing Centre, Forschungszentrum J\"ulich, D-52428 J\"ulich, Germany}

\affiliation[c]{Institute for Theoretical Physics, E\"otv\"os University, H-1117 Budapest, Hungary}

\affiliation[d]{Department of Physics, Pennsylvania State University, University Park, PA 16802, USA}

\emailAdd{a.kotov@fz-juelich.de}

\abstract{
We discuss properties of Quantum Chromodynamics at finite temperature obtained by means of lattice simulations with overlap fermions. This fermion discretization preserves chiral symmetry even at finite lattice spacing. We present details of the lattice formulation, first results for the chiral observables and discuss the behaviour of the system near the chiral thermal phase transition. 
}

\FullConference{The 40th International Symposium on Lattice Field Theory (Lattice 2023)\\
July 31st - August 4th, 2023\\
Fermi National Accelerator Laboratory\\}

\begin{document}
\maketitle

\section{Introduction}

One of the most interesting results obtained by means of Lattice Quantum Chromodynamics is the nature of the chiral phase transition at finite temperature, which turned out to be a crossover, rather than a genuine phase transition \cite{Aoki:2006we}. The major shortcoming of the \cite{Aoki:2006we} and subsequent studies of the thermal QCD is that they are mainly performed with staggered fermions, which 
possess only part of the chiral symmetry at finite lattice spacing, while the full symmetry group is expected to be recovered only after a not always simple continuum extrapolation. In this proceeding we discuss our current study of the chiral phase transition in QCD with another fermion discretization, namely overlap fermions \cite{Neuberger:1997fp}, which have the exact chiral symmetry already at finite lattice spacing. We present and discuss the first results for the chiral observables and the spectrum of the Dirac operator.

\section{Lattice details}

Our study is based on the same lattice setup, which was used in the previous studies with heavier pion masses \cite{Borsanyi:2012xf,Borsanyi:2015zva}. Now we extent the simulation details to $N_f=2+1$ light quark flavours and physical pion mass. The overlap implementation of odd number of flavours, which is used for the strange quark sector, was introduced in \cite{Borsanyi:2016ksw}.
We summarize the details of the action below:

\begin{itemize}
    \item tree-level Symanzik improved gauge action
    \item two light quark flavours with the action given by overlap Dirac operator:
    \begin{equation}
    aD_{\mathrm{ov}}=\frac{1}{2}\left(1+\gamma_5 \mathrm{sign}(\gamma_5 D_W(-m_W))\right),
    \label{eq:overlapaction}
    \end{equation}
    for the kernel we used Wilson Dirac operator $D_W(-m_W)$ with negative mass $-m_W=-1.3$
    \item one strange quark flavour of the overlap Dirac operator with the same action (\ref{eq:overlapaction})
    \item two fermion flavours with Wilson action, given by $D_W(-m_W)$. They are irrelevant in the continuum limit
    \item two boson fields (also irrelevant in the continuum limit) with the mass $m_B=0.54$ and the action:
    \begin{equation}
        \phi^{\dag}\left[D_W(-m_W)+im_B\gamma_5\tau_3\right]\phi
    \end{equation}
\end{itemize}

\begin{table}[thb]
    \centering
    \begin{tabular}{c|c|c}
    $N_t$ & $N_s$ & Aspect ratio $N_s/N_t$\\
    \hline
     8   & 16, 24, 32 & 2, 3, 4  \\
     10   & 20, 30, 40 & 2, 3, 4  \\
     12   & 24, 36, 48 & 2, 3, 4  \\
    \end{tabular}
    \caption{Summary of lattice sizes used in the current study}
    \label{tab:ensembles}
\end{table}

The main goal for the introduction of the auxiliary bosonic and fermion fields is to suppress the tunnellings between different topological sectors \cite{Fukaya:2006vs}. In this study we restrict simulations to the zero topological sector $Q=0$. Simulations in the fixed topological sector lead to finite volume effects \cite{Fukaya:2006vs}, which disappear in the limit $V\to\infty$. 

Generation of the configurations was done with the help of the standard Hybrid Monte-Carlo algorithm. For the application of the overlap Dirac operator one needs  to calculate the sign function of a hermitian Wilson-Dirac operator $X=\gamma_5D_W(-m_W)$: $\mathrm{sign}X = \frac{X}{\sqrt{XX^{\dag}}}$. In order to calculate it we first computed 32-128 low-lying modes of operator $X$, on which the sign function can be calculated explicitly, and for the rest of the spectrum we used Chebyshev polynomial approximation of the sign function. For the calculation of eigenvalues we used implicitly restarted Lanczos algorithm and the inversion of the overlap Dirac operator was done using flexible GMRES method with Wilson-Dirac operator as preconditioner. 

We performed numerical calculations along the line of constant physics (LCP), which was determined from calculations with larger pion masses in \cite{Borsanyi:2016ksw}. In Tab.~\ref{tab:ensembles} we summarize the ensembles which we use in our study. For each value of $N_t$ and $N_s$ we performed simulations for eight temperatures: $T=140, 145, 150, 155, 160, 165, 170, 175$ MeV.

\begin{figure}[htb]
    \centering
    \includegraphics{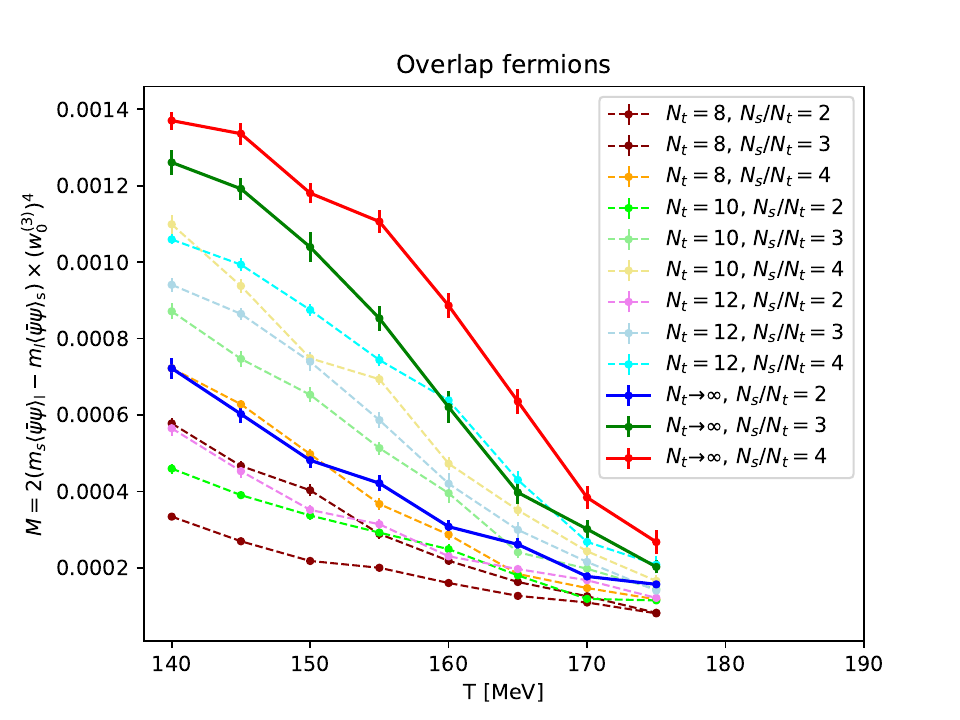}
    \caption{Chiral condensate as a function of temperature for all lattice sizes used in the current study. Solid lines correspond to the continuum extrapolation at fixed aspect ratio $N_s/N_t$.}
    \label{fig:chcond}
\end{figure}

\section{Chiral phase transition}

For the study of the chiral phase transition we measured the chiral condensate and the chiral susceptibility. Lattice data at finite lattice spacing have to be renormalized. In this study we employed the renormalization procedure, based on the strange quark condensate. The renormalized observables are defined then:

\begin{equation}
\begin{split}
    \langle \bar{\psi}\psi \rangle_r &= 2(m_s\langle \bar{\psi}\psi \rangle_l-m_l\langle \bar{\psi}\psi \rangle_s)\\
    \chi_r &= m_s\partial_{m_l} M
    \label{eq:chcondsus}
\end{split}
\end{equation}

This renormalization procedure  does not eliminate all additive divergences, logarithmic terms $\sim m^3\log ma$ remain in Eq. (\ref{eq:chcondsus}), however their numerical values is expected to be small due to suppression by the quark mass \cite{HotQCD:2019xnw}.

\begin{figure}[ht]
    \centering
    \includegraphics[width=7.5cm]{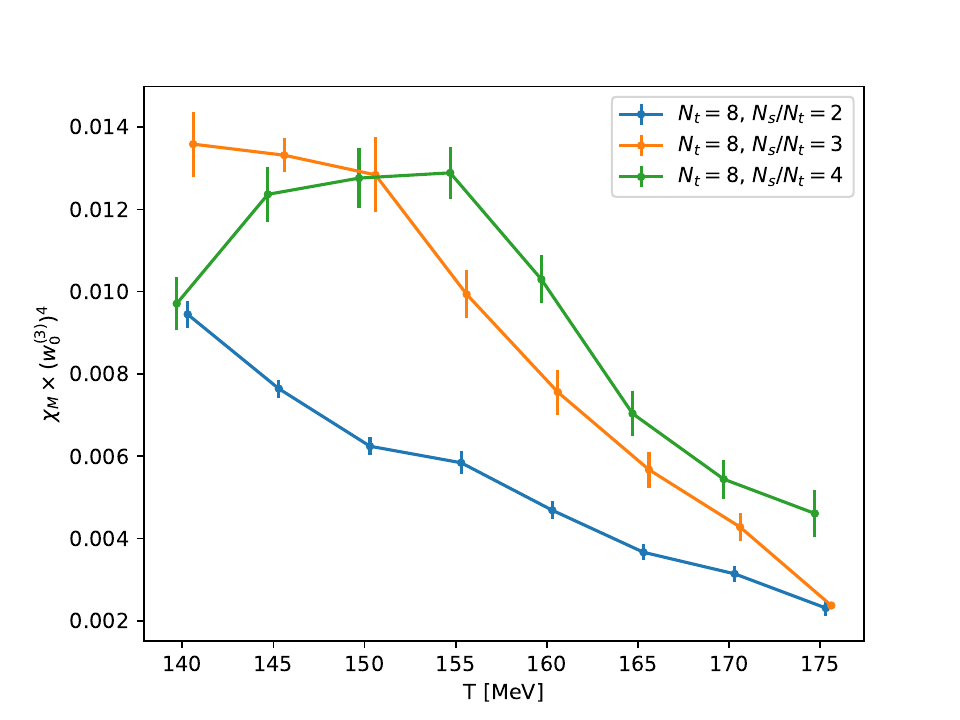}
    \includegraphics[width=7.5cm]{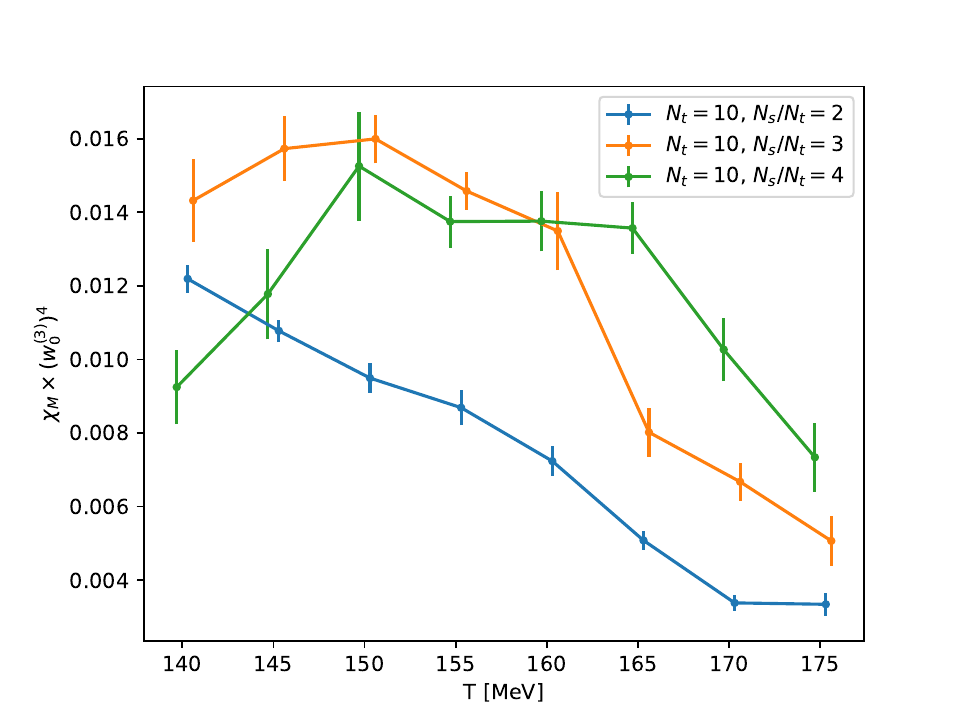}
    \includegraphics[width=7.5cm]{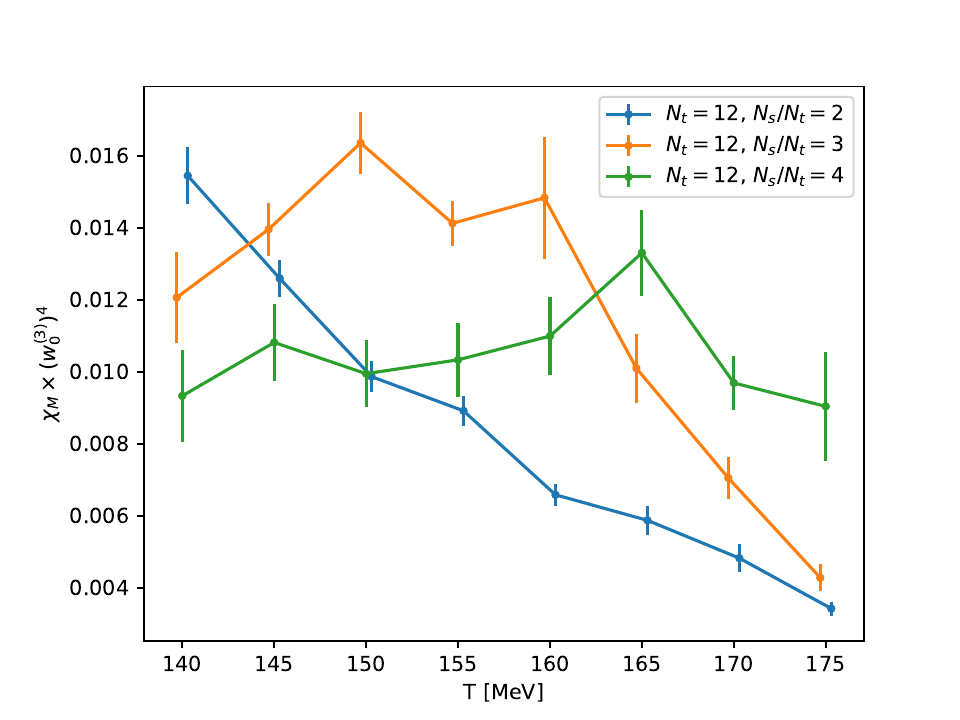}
    \caption{Chiral susceptibility as a function of temperature for all lattice sizes used in the current study..}
    \label{fig:chsus}
\end{figure}

In Fig.~\ref{fig:chcond} we present the chiral condensate for all  ensembles used in the current study, shown by dashed lines. One can clearly see that results significantly depend both on the lattice spacing and on the spatial volume. In addition, in Fig.~\ref{fig:chcond} we also show the continuum extrapolation  of the data for each volume, obtained with a simple $\sim a^2$ behaviour. The results also suggest large volume dependence, for aspect ratio $N_s/N_t=2$ there is no clear inflection point in the studied temperature range $T\in [140, 175]$ MeV, while larger aspect ratios $N_s/N_t=3,4$ exhibit an expected threshold behaviour with inflection point $T\sim 160$ MeV.

In Fig.~\ref{fig:chsus} we show the chiral susceptibility for all ensembles. The behaviour of the chiral susceptibility is consistent with the behaviour of the chiral condensate: for small aspect ratio $N_s/N_t=2$ there is no clear peak and consequently no indication of the phase transition $T\in [140, 175]$ MeV, while large aspect ratios $N_s/N_t=3,4$ exhibit a peak-like behaviour with pseudo-critical temperature $T\sim 160$ MeV. We would like to note that recent studies with the staggered fermions at finite volume \cite{Borsanyi:2024wuq} suggest that for aspect ratio $N_s/N_t=2$ the pseudo-critical temperature is smaller $T\sim140$ MeV than the pseudo-critical temperatue in the infinite volume $T\approx155$ MeV, which is in agreement with our results obtained with the overlap fermions. It is worth noting that our studies are done in the fixed topological sector $Q=0$, which is another source of finite volume effects and can also affect the results in the fixed volume. We plan to study how simulations in the fixed topological sector $Q=0$ affect our data in the future.

\begin{figure}[htp]
    \centering
    \includegraphics[width=7cm]{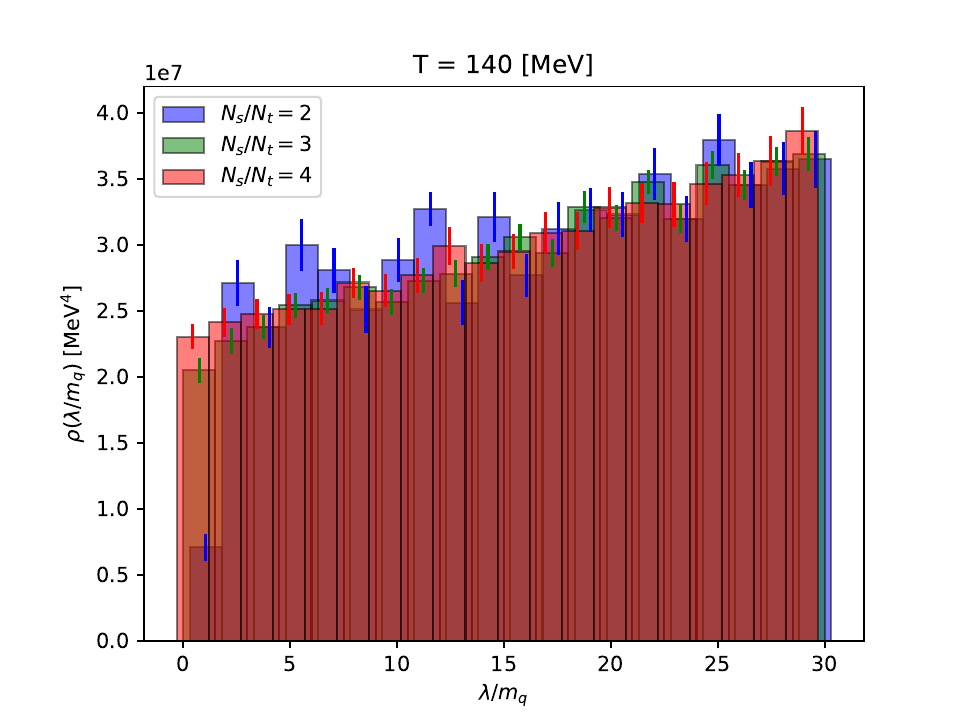}
    \includegraphics[width=7cm]{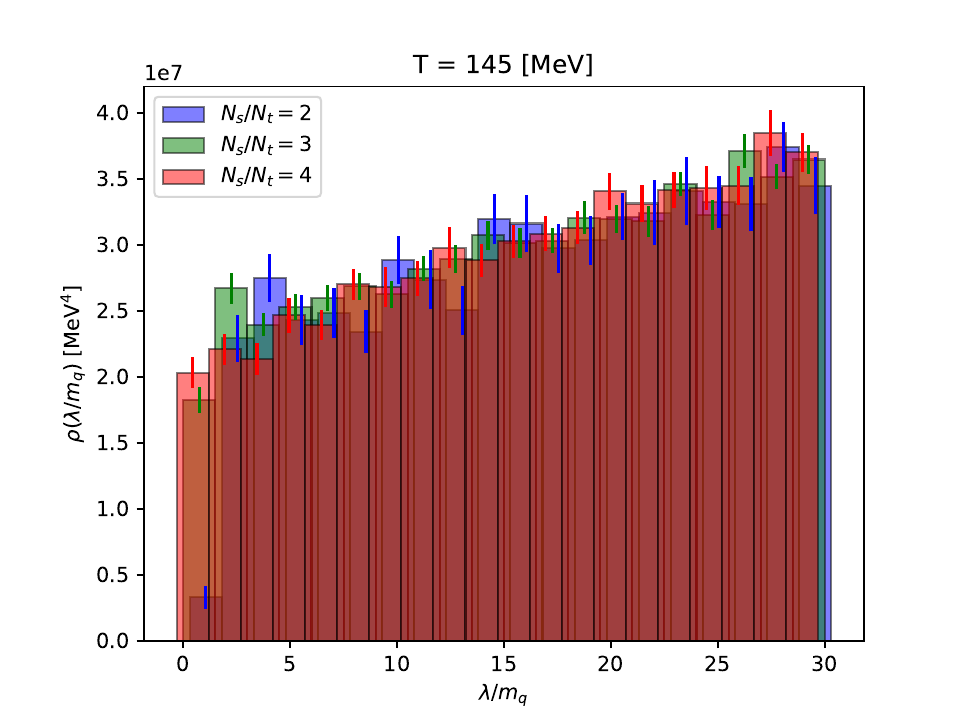}
    \includegraphics[width=7cm]{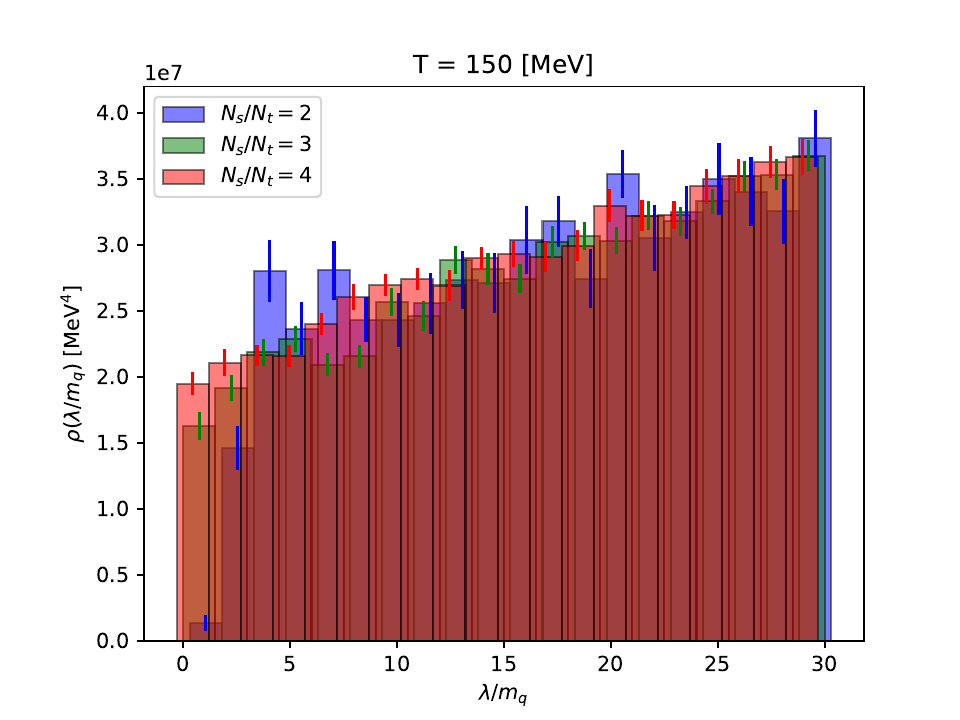}
    \includegraphics[width=7cm]{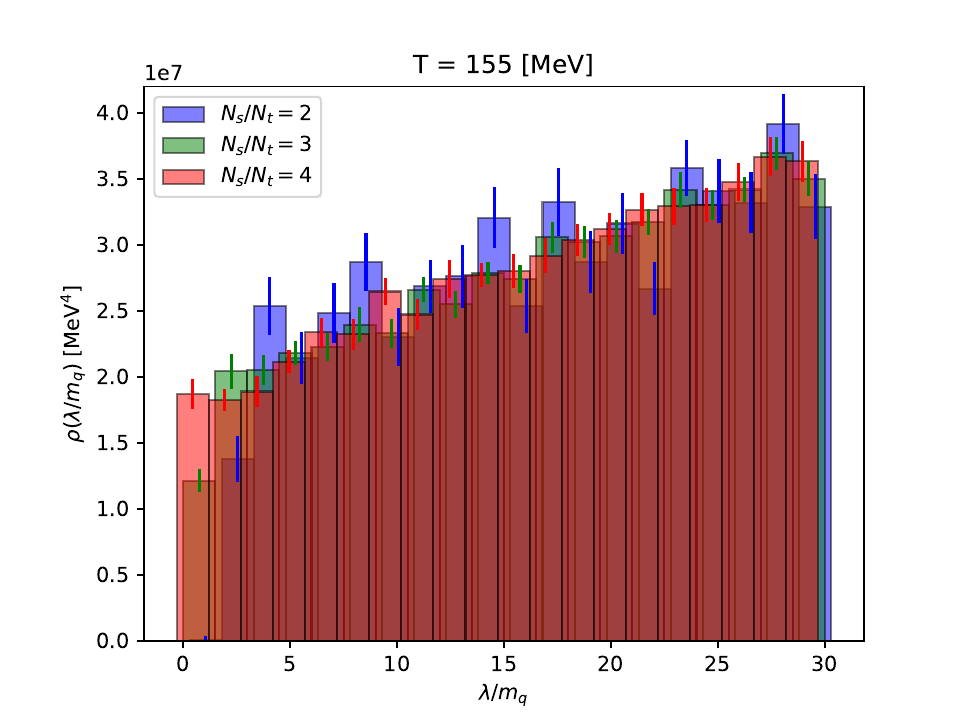}
    \includegraphics[width=7cm]{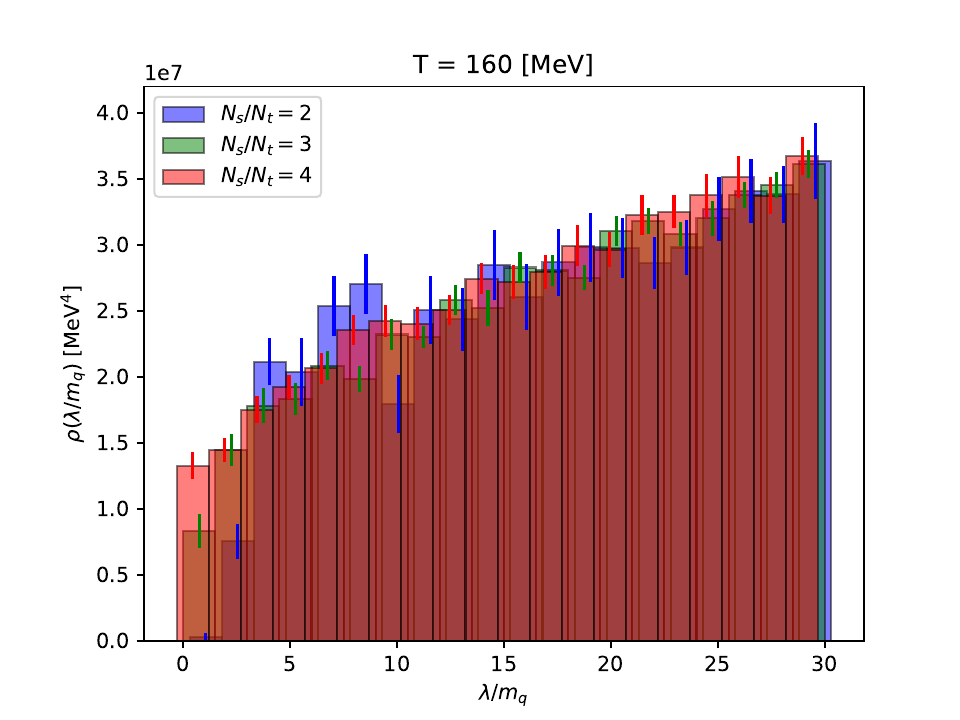}
    \includegraphics[width=7cm]{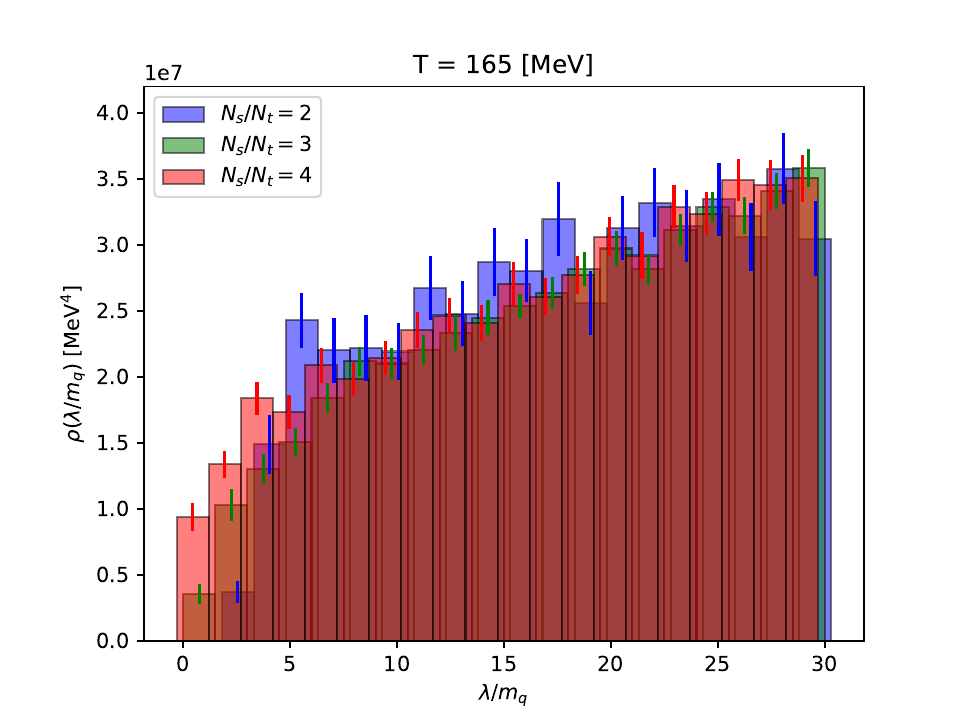}
    \includegraphics[width=7cm]{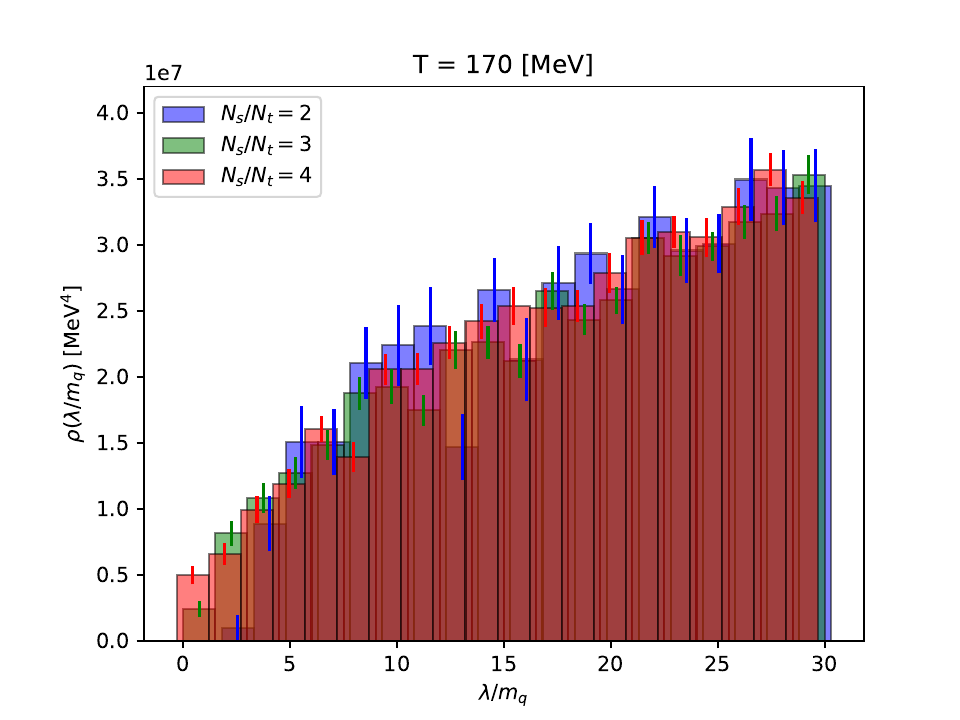}
    \includegraphics[width=7cm]{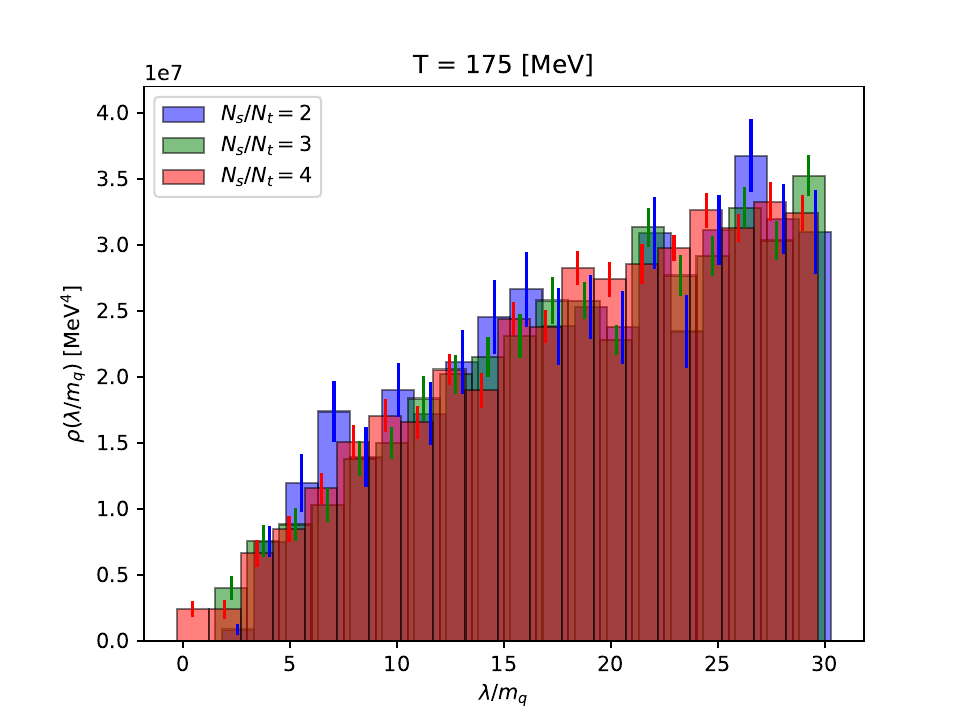}
    \caption{Continuum extrapolated spectrum of the Dirac operator $\rho(\lambda/m_q)$ versus $\lambda/m_q$ for aspect ratios $N_s/N_t=2$, $3$, $4$. }
    \label{fig:spectrum}
\end{figure}

\section{Spectrum of the Dirac operator}

Apart from the chiral condensate, we also studied the spectrum of the overlap Dirac operator, described by the spectral function $\rho(\lambda)$. It is related to the chiral condensate via Banks-Casher relation \cite{Banks:1979yr}:
\begin{equation}
    \langle\bar{\psi}\psi\rangle_r=2\int_0^{\infty} d\lambda\rho(\lambda)\frac{m}{m^2+\lambda^2}\xrightarrow[m\to 0]{} \pi \rho(\lambda).
\end{equation}

The behaviour of $\rho(\lambda)$ has been  subject of several studies \cite{Alexandru:2019gdm, Kehr:2023wrs, Kovacs:2023vzi}, which suggest rather nontrivial behaviour near origin with a formation of the peak, possibly indicating some new phase in QCD. In Fig.~\ref{fig:spectrum} we present the results for the continuum extrapolation of the spectrum for all values of $N_t$ and aspect ratios used in our spectrum. In order to avoid issues with the renormalization we used the ratio $\lambda/m_q$, in which renormalization constants are cancelled \cite{Giusti:2008vb}.

We observe that the results for the spectral density $\rho$ almost do not depend on the aspect ratio, except for the behaviour at small $\lambda$, where for the smallest studied aspect ratio the spectral density $\rho(\lambda/m_q)$ is suppressed as compared to larger aspect ratios. It can be attributed to the fact that low-lying large distance modes simply do not fit into small lattice volumes. For this reason we also do not observe standard signatures of the transition-like behaviour at small $N_s/N_t=2$. However, data at larger $N_s/N_t=3,4$ show a nonzero intercept as $\lambda\to0$ which disappears if one increases temperature at $T\approx160-165$ MeV. This behaviour is in a full agreement with the results extracted from the chiral condensate and the chiral susceptibility. Our data also do not suggest the existence of the large peak near origin for the temperatures and volumes under study. 

\section{Conclusions}

We presented the first results of our study of the chiral QCD phase transition with dynamical overlap fermions. This fermion discretization possesses chiral symmetry at finite lattice spacing. We restricted ourselves to the simulations in the fixed toplogical sector $Q=0$. We showed the dependence of the chiral condensate and the chiral susceptibility on the temperature in the vicinity of the chiral phase transition $T\in[140,175]$ MeV for three values of $N_t=8,10$ and $12$ and three spatial volumes corresponding to aspect ratios $N_s/N_t=2,3,4$. Our results for the smallest aspect ratio $N_s/N_t=2$ most likely are subject to large finite volume effects and do not suggest the existence of the phase transition or crossover in the temperature range $T\in[140,175]$ MeV, while for two larger aspect ratios $N_s/N_t=3,4$ the pseudo-critical temperature $T\approx160$ MeV is in the agreement with previous staggered studies. We also presented the results for the spectrum of the Dirac operator, which is consistent with the same picture and the same estimation of the pseudo-critical temperature.

\section{Acknowledgements}
The authors gratefully acknowledge computing time on the supercomputer JURECA\cite{JURECA} at Forschungszentrum Jülich under grant no. qcdoverlap.

\bibliographystyle{unsrt}
\bibliography{overlap}

\end{document}